\documentclass[a4paper]{iopart}
\usepackage{latexsym}
\usepackage[dvips]{graphicx}   
\usepackage[all]{xy}           
\usepackage{amsmath}           
\usepackage{amsfonts}          
\usepackage{amssymb}           

\usepackage[english]{babel}

\newcommand{\ef}[1]{\, #1}     

\newcommand{\references}{
\section*{References}}

\newcommand{\ham}{\mathcal{H}}

\newcommand{\eval}[1]{\left\langle {#1} \right\rangle}

\newcommand{\reval}[1]{\overline{#1}}

\newcommand{\dx}[1] {\mathrm{d}{#1}}
\newcommand{\dede}[1]{\frac{\partial}{\partial #1}}

\newcommand{\vett}[1]{#1}

\newcommand{\grid}{J}

\newtheorem{theor}{Theorem}

\begin{document}

\jl{1}

\def\PACS{\par\leavevmode\hbox {\it PACS:05.50.+q, 64.60.Cn, 75.10.-b}}

\article
{}
{General duality for abelian-group-valued 
statistical-mechanics models}
\author
[
  Sergio Caracciolo,
  Andrea Sportiello
]
{
  Sergio Caracciolo$^{[1]}$,
  Andrea Sportiello$^{[2]}$
}
\address{[1] Universit\`a degli Studi di Milano - Dip. di Fisica and INFN,
\\ via Celoria 16, I-20133 Milano,  and NEST-INFM, Italy
\\ Mail address: \tt{Sergio.Caracciolo@mi.infn.it}}
\address{[2] Scuola Normale Superiore and INFN - Sez. di Pisa,
\\ Piazza dei Cavalieri 7, I-56100 Pisa, Italy
\\ Mail address: \tt{Andrea.Sportiello@sns.it}}

\date{August 25$^{th}$, 2003}
\begin{abstract}
We introduce a general class of statistical-mechanics models, 
taking values in an abelian group, which includes
examples of both spin and gauge models, both ordered and disordered.

The model is described by a set of ``variables'' 
and a set of ``interactions''.
Each interaction is associated with a linear combination of variables; these
are summarized in a matrix $J$.
A Gibbs
factor is associated to each variable (one-body term) and to each
interaction.

Then we introduce a duality transformation
for systems in this class.
The duality exchanges the abelian group with its dual, the
Gibbs factors
with their Fourier transforms, and the interactions with the variables.
High (low) couplings in the interaction terms are mapped into
low (high) couplings in the one-body terms. 
If the matrix $J$ is interpreted as a
vector representation of a matroid, duality 
exchanges the matroid with its dual.

We discuss some physical examples. The main idea is to generalize
known models up to eventually include randomness into the pattern of
interaction.
We introduce and study a random Gaussian Model, a random
Potts-like model, and a random variant of discrete scalar QED.

Although the classical procedure {\em 'a la} Kramers and
Wannier does not extend in a natural way to such a wider class of
systems, our weaker procedure applies to these models, too.
We shortly describe the consequence of duality for each example.
\end{abstract}

\PACS

\noindent
{\it keywords: Duality, Matroid Theory, Potts model.
}


\section{Introduction of the problem}

In statistical mechanics a notion of duality appeared first 
as a relation between
the high- and low-temperature expansions in the two-dimensional Ising
model~\cite{kw}. 
In both expansions the partition function can be expressed as a sum on a gas of
closed polymers: in the low-temperature picture, these polymers are the
boundaries of the ferromagnetic domains, so we have an expansion in the
deviations from the ordered regime;
on the contrary, in the high-temperature picture
the polymers are the terms of a Fourier 
expansion which expresses the deviation from a
disordered regime of total decorrelation. 
So, in a certain sense, the duality relates {\em energetic} degrees of freedom
to {\em entropic} degrees of freedom.
Moreover, under the assumption of the existence of only one
critical value for the temperature, it was possible to locate the
phase-transition point before the exact solution of the model by Onsager.

More formally, a duality transformation can be defined directly on the
partition
function of the model by means of two independent duality transformations,
that is:
\begin{itemize}
\item The variables are located on the
  $k$-cells of a $D$ dimensional cell-complex 
  (e.g.~the $0$-cells, or {\em sites}, for a spin model, the $1$-cells, or
  {\em bonds}, for a gauge model), more precisely the physical state
  is a $k$-cochain taking values in an abelian group. 
  The interactions are located on the $(k+1)$-cells (respectively the
  {\em bonds} and the {\em plaquettes} in the previous examples). 
  The ``algebraic'' duality exchanges variables and
  interactions, and the dual physical state is a
 $(k+1)$-chain on the dual of the original group.
\item The new variables have a gauge redundancy, and must obey a set of
  constraints: in other words, they must be a closed $(k+1)$-chain. 
  The solution
  of the constraints leads to the ``integration'' of the $(k+1)$-chain to a
  $(k+2)$-chain. After this, the cell-complex 
  on which the model is defined is transformed into its
  dual counterpart.
  This ``geometrical'' duality of cellular complexes pictorially exchanges
  $k$-cells with $(D-k)$-cells, and the physical state is now a
  $(D-k-2)$-cochain.
\end{itemize}
This is the general strategy, explained in detail, for example, in the review
part of \cite{druhlwagner}, or in \cite{savit}.

In this paper we wish to clarify the first (``algebraic'') part of the duality
transformation, by putting it in a more general context. We consider a
rather general class of statistical-mechanical models in which the variables
take values in an abelian group; each interaction term is associated to a
particular linear combination of variables, but need not arise from any
particular geometrical entity (such as a bond or a plaquette).
By relaxing the requirement for a cell-complex structure,
we can deal with a wider class of statistical models.
We think that this more general notion of duality can be fruitful
in the study of disordered systems. 

This notion of duality is motivated, in fact, by the 
{\em clauses-variables} duality 
that we observed in a previous investigation in
the zero-temperature behaviour of a model of random 
satisfiability~\cite{us}, which has the structure of a 
disordered diluted interaction.


The paper is organized as follows:
\begin{itemize}
\item
In section 2 we introduce some mathematical
formalism, in particular about the notion of duality and Fourier transform
for generic abelian groups.
\item
In the sections from 3 to 8 we discuss our
duality under general aspects: in section 3 we describe the most general
setting for the duality; 
in section 4 we specialize the model to a formulation which is manifestly
suitable for statistical mechanics models, and in section 5 we
rederive the duality in this frame; 
in section 6 we show how to
reconduct an extension of the original definition to the primitive
formulation of the problem, which is reminiscent of
a set of gauge fields coupled to the interaction terms; 
in section 7 we describe a particular extra-feature appearing in the
model when delta constraints appears;
finally in section 8 we introduce a sufficient criterium for self-duality.
\item
In sections from 9 to 13 we show the application of these general ideas to
some concrete physical examples: first, in sections 9 and 10, 
a sort of random gaussian
model, and a variant with random constraints; further, in sections
11 and 12, a generalization
of Potts model extended to random structures of interactions, and a
discussion on how to recover, inside our framework,
the traditional results for regular lattices; 
finally, in section 13, we discuss a
concrete example of the general setting of section 6, which is
reminiscent of a scalar field theory coupled to electro-magnetism,
both in the case of regular lattices and of disordered systems.
\item
In section 14 we outline some conclusions, and introduce the perspective for
a further generalization of the procedure, using different mathematical
structures instead of an abelian group structure and the Fourier transform.
\end{itemize}

\section{Dual for an abelian group, Fourier transform 
and group homomorphisms}

In this section we introduce the {\em Fourier transform} 
for locally compact abelian groups, 
by means of the theory of characters, following
the textbook of Kirillov and Gvishiani \cite{kiril}.

First, we introduce the concept of {\em dual group}.
Let $G$ be a locally compact abelian group (we always write our abelian groups
additively). Let $\mathcal{R}_G$ be the set of irreducible representations
of $G$. It is well known that the irreducible representations of $G$ are all
one-dimensional: thus, for each $\rho \in \mathcal{R}_G$,
the representation matrix 
$U_{\rho}(g)$,
is one-dimensional and
coincides with the character $\chi_{\rho}(g)$.
The set $\mathcal{R}_G$ is
itself an abelian group under ordinary composition (multiplication); we write
this group additively and call it the dual group
$\widehat{G}$.
It is understood in the following that latin letters 
$g$, $h$\ldots~are elements of the original group $G$, 
and greek letters like 
$\rho$, $\tau$\ldots~are elements of the dual group $\widehat{G}$.
If we adopt the notation
\begin{equation}
\chi_{\rho}(g) = U_{\rho}(g) = e^{i \vartheta_G(\rho,g)}
\ef{,}
\end{equation}
the following bilinear relations hold:
\begin{subequations}
\label{group}
\begin{gather}
\label{thetanot}
\vartheta_G(\rho + \tau, g)= \vartheta_G(\rho,g) + \vartheta_G(\tau,g)
\ef{;}
\\
\vartheta_G(\rho, g+h)= \vartheta_G(\rho,g) + \vartheta_G(\rho,h)
\ef{.}
\end{gather}
\end{subequations}
It can be shown that the dual group of $\widehat{G}$, 
denoted~
\makebox{\raisebox{4pt}[0pt][4pt]{$\widehat{}$}%
\raisebox{2.5pt}[0pt][4pt]{$\widehat{}$}%
\!\!\!$G$}, is naturally isomorphic to $G$.

In a consistent choice of notations, 
the characters of the dual group are
derivable from the antisymmetric relation
\begin{equation}
\vartheta_{G}(\rho, g) = 
-\vartheta_{\widehat{G}}(g, \rho) 
\ef,
\end{equation}
which is equivalent to
\begin{equation}
\chi_{g}(\rho)=\reval{\chi_{\rho}(g)}
\ef.
\end{equation}
Given a function $f$ from $G$ to the 
complex plane, its Fourier transform $\widehat{f}$ is a function from the dual
group $\widehat{G}$ to the complex plane, defined by
\begin{subequations}
\begin{align}
f:&\ G \rightarrow \mathbb{C}
&
f(g) &=
\int_{\rho}^*
e^{i \vartheta_{\widehat{G}}(g, \rho)} \widehat{f}(\rho)
      \\
\widehat{f}:&\ \widehat{G} \rightarrow \mathbb{C}
&
\widehat{f}(\rho) &=
\int_g e^{i \vartheta_{G}(\rho, g)} f(g)
\ef{.}
\end{align}
\end{subequations}
where $\int_g$ and $\int_{\rho}^*$ stand 
for a properly normalized Haar measure on $G$ and $\widehat{G}$ (see
below). Subscripts on the bilinear forms 
$\vartheta_{G}$ and $\vartheta_{\widehat{G}}$ will be
omitted when clear.

\begin{table}[tb!]
\[
\begin{array}{|r|c|c|c|}
\cline{2-4}
\multicolumn{1}{r|}{}
& \rule[-5mm]{0pt}{11mm}
\rule{5mm}{0pt}
 \displaystyle{(\mathbb{Z}_q ; \mathbb{Z}_q)} 
\rule{5mm}{0pt}
&
\rule{4mm}{0pt}
 \displaystyle{(U(1) ; \mathbb{Z})}
\rule{4mm}{0pt}
&
\rule{6mm}{0pt}
 \displaystyle{(\mathbb{R} ; \mathbb{R})}
\rule{6mm}{0pt}
\\
\hline
\rule[-7mm]{0pt}{16mm}
\rule{8mm}{0pt}
\displaystyle{\int_{g} :}  
& \displaystyle{\frac{1}{q} \sum_{x=0}^{q-1}} 
& \displaystyle{\int_0^{2 \pi} \frac{\mathrm{d}\theta}{2 \pi}}
& \displaystyle{\int_{-\infty}^{\infty} \dx{x}} 
\\
\rule[-6mm]{0pt}{14mm} 
\displaystyle{\int_{\rho}^* :}  
& \displaystyle{\sum_{\xi=0}^{q-1}}
& \displaystyle{\sum_{n=-\infty}^{+\infty}}
& \displaystyle{\int_{-\infty}^{\infty} \dx{\xi}} 
\\
\rule[-3mm]{0pt}{9mm} 
\displaystyle{\vartheta(g, \rho) :}  
& \displaystyle{2 \pi x \xi /q}
& \displaystyle{\theta n}
& \displaystyle{2 \pi x \xi}
\\
\rule[-3mm]{0pt}{9mm} 
\displaystyle{\delta_G(g) :}  
& \displaystyle{q \,\delta_{\textrm{Kr}}(x)}
& \displaystyle{2 \pi \delta_{\textrm{Dirac}}(\theta)}
& \displaystyle{\delta_{\textrm{Dirac}}(x)}
\\
\rule[-5mm]{0pt}{11mm} 
\displaystyle{\delta_{\widehat{G}}(\rho) :}  
& \displaystyle{\delta_{\textrm{Kr}}(\xi)}
& \displaystyle{\delta_{\textrm{Kr}}(n)}
& \displaystyle{\delta_{\textrm{Dirac}}(\xi)}
\\
\hline
\end{array}
\]
\caption{Convention for the normalization of the Haar measure 
  adopted in this paper
  for various pairs $(G, \widehat{G})$ of abelian groups.
  It is understood that for the direct
  product of groups the Haar measure is simply the product measure of the
  original ones.}\label{t1}
\end{table}
We introduce a notion of invariant $\delta$-function compatible with the
measure introduced above:
\begin{align*}
\int_{\rho}^* \overline{\chi_{\rho}(g')} \ \chi_{\rho}(g) &= 
\delta_G(g'-g)
\ef{;}
&
\int_{g} \overline{\chi_{\rho'}(g)} \ \chi_{\rho}(g) &= 
\delta_{\widehat{G}}(\rho'-\rho)
\ef{;}
\end{align*}
which have the natural property
\begin{align*}
\int_{g} f(g) \ \delta_G(g'-g) &= 
f(g')
\ef{;}
&
\int_{\rho}^* \widehat{f}(\rho) \ \delta_{\widehat{G}}(\rho'-\rho) &= 
\widehat{f}(\rho')
\ef{.}
\end{align*}
Inside the wide class of ``physically natural'' groups with the
properties of being locally compact and compactly generated,
the most general abelian group is 
the product of $\mathbb{R}$, $U(1)$, 
$\mathbb{Z}$ and $\mathbb{Z}_q$ groups,
so the suggested choice of normalizations
of Table~\ref{t1}
covers all the possible situations.
\footnote{This fact derives essentially from three general results
\cite{HR}:
\begin{itemize}
\item Every locally compact, compactly generated abelian group $G$ is
topologically isomorphic with 
$\mathbb{R}^{a} \times \mathbb{Z}^{b} \times K$ for some nonnegative
integers $a$ and $b$ and some compact abelian group $K$.
\item Let $K$ be a compact abelian group and let $U$ be a neighborhood
of the identity in $K$. There is a closed subgroup $H$ of $G$ such
that $H \subset U$ and $K/H$ is topologically isomorphic with
$U(1)^{c} \times F$, where $c$ is a nonnegative integer and $F$ is a
finite abelian group.
\item Let $F$ be a finite abelian group. Then it is isomorphic to the
group direct product of cyclic groups of prime power order
({\em Kronecker decomposition theorem}).
\end{itemize}}




A homomorphism $\varphi$ 
from an abelian group $G_1$ to another abelian group $G_2$
is a map which preserves the group
operation $\varphi(x_1 +_{G_1} x_2)= \varphi(x_1) +_{G_2} \varphi(x_2)$, and
maps the identity to the identity, $\varphi(0)=0$.
We denote by
$\mathit{Hom}(G_1, G_2)$ the set of all homomorphisms from $G_1$ to $G_2$.
Note that a homomorphism
must preserve additive inverses, i.e.~$\varphi(-g)=-\varphi(g)$. This follows
from $\varphi(g)+\varphi(-g)=\varphi(0)=0$.

As we have seen in the previous paragraphs,
from theory of characters we can introduce the dual
groups $\widehat{G}_1$ and $\widehat{G}_2$.
Following Kirillov (\cite{kiril}, Part III, statement 631),
it is indeed possible to introduce a notion of {\em dual homomorphism}
$\varphi^T \in \mathit{Hom}(\widehat{G}_2,\widehat{G}_1)$ 
which completes the diagram
\begin{equation*}
\xymatrix{
G_1
\ar@{<-->}[rr]^{\mathcal{D}}
\ar[d]_{\varphi}
&&
\widehat{G}_1
\\
G_2
\ar@{<-->}[rr]^{\mathcal{D}}
&&
\widehat{G}_2
\ar[u]_{\varphi^T}
}
\end{equation*}
The dual homomorphism $\varphi^T$ acts according to the formula
\begin{equation}
\label{dualhom}
\chi_{\rho}(\varphi(g))
=\chi_{g}(\varphi^T(\rho))
=\reval{\chi_{\varphi^T(\rho)}(g)}
\ef;
\qquad
g \in G_1
\ef;
\quad
\rho \in \widehat{G}_2
\ef,
\end{equation}
from which we derive
\begin{equation}
\vartheta_{G_2}(\varphi(x),y)=\vartheta_{G_1}(\varphi^T(y),x)
\ef.
\end{equation}
It is easy to verify that $\varphi^T$ is a homomorphism from 
$\widehat{G}_2$ to $\widehat{G}_1$, as a consequence of
the fact that $\vartheta(\rho, g)$ is a bilinear form~\footnote{The
homomorphism property descends from the bilinearity properties 
(\ref{group}):
\begin{equation*}
\chi_{\varphi^T(\rho_1 + \rho_2)}(g)
=
\reval{\chi_{\rho_1 + \rho_2}(\varphi(g))}
=
\reval{\chi_{\rho_1}(\varphi(g))}
\cdot
\reval{\chi_{\rho_2}(\varphi(g))}
=
\chi_{\varphi^T(\rho_1)}(g)
\cdot \chi_{\varphi^T(\rho_2)}(g)
=
\chi_{\varphi^T(\rho_1)+ \varphi^T(\rho_2)}(g)
\ef.
\end{equation*}
}.


\section{The general formulation of the duality}
\label{gensec}

In this section we wish to 
introduce a notion of duality for a wide class of systems in classical
statistical mechanics.

In the most general form, given two abelian groups $G_1$ and $G_2$,
we will consider the system defined by
\begin{itemize}
\item a function $F_1 : G_1 \rightarrow \mathbb{C}$;
\item a function $\widehat{F}_2 : \widehat{G}_2 \rightarrow \mathbb{C}$;
\item a homomorphism $\varphi \in \mathit{Hom}(G_1,\widehat{G}_2)$;
\end{itemize}
and the Gibbs weight of a state $x \in G_1$ is given by
\begin{equation}
\label{ham1p}
\exp \left[ -\ham_{F_1,\widehat{F}_2 ;\varphi}(x) \right]
=
F_1(x) \widehat{F}_2(\varphi(x))
\ef.
\end{equation}
The partition function is defined as
\begin{equation}
Z_{F_1,\widehat{F}_2;\varphi}
=
\int_{x}
F_1(x) \widehat{F}_2(\varphi(x))
\ef.
\end{equation}
Applying Fourier transform to the function $\widehat{F}_2$ 
we obtain an algebraic relation:
\begin{multline*}
\int_{x}
F_1(x) \widehat{F}_2(\varphi(x))
=
\int_{x} \int_{y}
F_1(x) F_2 (y) e^{i \vartheta_{G_2}(\varphi(x),y)}
\\
=
\int_{y}
\int_{x} 
F_2 (y) 
F_1 (x) 
e^{i \vartheta_{G_1}(\varphi^T (y),x)}
=
\int_{y}
F_2 (y) 
\widehat{F}_1 (\varphi^T(y)) 
\ef,
\end{multline*}
from which we deduce that
\begin{equation}
\label{dual1p}
Z_{F_1,\widehat{F}_2;\varphi}
=
Z_{F_2,\widehat{F}_1;\varphi^T}
\ef.
\end{equation}
Note that in the dual model the roles of the group $G_1$ and $G_2$ are
interchanged, and the homomorphism 
$\varphi \in \mathit{Hom}(G_1,\widehat{G}_2)$ is replaced by
$\varphi^T \in \mathit{Hom}(G_2,\widehat{G}_1)$.

\section{Definition of a model with many degrees of freedom}
\label{prodsec}

In this section we specialize the procedure
described above to
a more physical frame, in which {\em extensive} parameters 
appear both in the choice of the groups $G_1$ and $\widehat{G}_2$ 
and of the functions
$F_1$ and $\widehat{F}_2$.

We introduce two extensive parameters: the first one,
$r$, is called {\em the number of variables}, while the second one, $(n-r)$,
is {\em the number of interactions} 
(they were called {\em clauses} in \cite{us}).

Given a group $G$, we adopt 
$G_1=G^{\otimes r}$ and $\widehat{G}_2=G^{\otimes (n-r)}$.
A state in $G_1$ is specified by a vector $x=(x_1, \ldots, x_r)$, and the
homomorphism $\varphi$ is specified by a $r \times (n-r)$ matrix $J$, with
elements $J_{ik}\in \mathit{Hom}(G,G)$ acting on the left, such that $(xJ)_k$
means
$\sum_{i=1}^{r} J_{ik}(x_i)$.
By convention we let the indices $k$ run from $r+1$ to $n$.

The functions $F_1$ and $\widehat{F}_2$ will be chosen to be
factorized on the entries of the vectors:
\begin{equation}
F_1[x]= \prod_{i=1}^{r} f_i(x_i)
\ef;
\qquad
\widehat{F}_2[y]= \prod_{k=r+1}^{n} f_k(y_k)
\ef.
\end{equation}
The weight functions $f_1$, \ldots $f_r$ are thus ``one-body terms''
(generalized magnetic fields), while the 
$f_{r+1}$, \ldots $f_n$ are ``interaction terms''.

Note that we have chosen
the groups $G_1$ and $\widehat{G}_2$ to be
direct products of {\em the same} group $G$. To avoid confusions
in the application of the conventions of Table~\ref{t1}, we specify that
\begin{align*}
\int_{[x]\in G_1} &\equiv \int_{x_1} \cdots \int_{x_r}
\ef;
&
\vartheta_{G_1}(x',x) &=
\sum_i \vartheta_{G} (x'_i, x_i)
\ef;
\\
\int^*_{[y]\in G_2} &\equiv \int^*_{y_{r+1}} \cdots \int^*_{y_n}
\ef;
&
\vartheta_{G_2}(y',y) &=
- \sum_k \vartheta_{G} (y_k, y'_k)
\ef.
\end{align*}
We will sometime use a notation in
which the two functions $F_1[x]$ and $\widehat{F}_2[xJ]$ 
are ``merged'' into only one
function $F[xB]=\prod_{j=1}^{n} f_j((xB)_j)$.

Denote with 1
the identity homomorphism $\varphi_I \in \mathit{Hom}(G,G)$, such that
$\varphi_I(x)=x$ for any $x$, and with 0 the null
homomorphism $\varphi_0 \in \mathit{Hom}(G,G)$, such that
$\varphi_0(x)=0$ for any $x$.
The $r \times n$ matrix $B$ is thus defined as
\begin{equation}
B_{ij}=
\left\{
\begin{array}{ll}
1 & i=j; \quad 1 \leq j \leq r;
\\
0  & i \neq j; \quad 1 \leq j \leq r;
\\
J_{ij} & r+1 \leq j \leq n;
\end{array}
\right.
\end{equation}
which can be written simply as
\begin{equation}
B=
\bigg[ \ I_r \  \bigg| \ J \  \bigg]
\ef{.}
\end{equation}
The invariance of the partition function under change of variables in the
integration corresponds
in terms of the matrix $B$ to the application of a special
orthogonal transformation from the left ({\em change of basis})
\begin{equation}
Z[B]=\int_{G^r} F[xB] \ef;
\qquad
Z[B]=Z[OB] \quad \forall \ O \in SO(r)
\ef.
\end{equation}
It results that, beyond the particular choice of the matrix $B$, 
a more fundamental notion of ``pattern of interaction'' 
can be established, which
is independent from the choice of basis.
This is the analogous of what happens in linear algebra, where the
notion of linear independence is intrinsic with respect to 
the choice of basis. 
The introduction of this concept in
mathematics dates up to the paper of Whitney \cite{whitney35}, and is the
foundation of a theory called {\em Matroid Theory} (cfr.~\cite{oxley} for a
textbook). In particular, {\em matroid} is the name given to the intrinsic
mathematical structure corresponding to our pattern of interaction, and a
choice for the matrix $B$ is called a {\em vectorial representation} for the
matroid. In the case in which the first $r \times r$ block of the matrix is
the identity, we say that $B$ is a {\em standard vectorial representation} for
the matroid. Note that our particular choice puts $B$ automatically in
standard vectorial form.

In this frame, we can develop a physical intuition on the structure of our
model. We
could consider the $r$ entries of the vector $x$ as the variables of a
statistical mechanic system.
The first $r$ functions $f_i(x_i)$ collect
the one-body contributions to the Gibbs factor, like one-body 
measure terms, or magnetic fields, or chemical potentials.
The last $(n-r)$ functions $f_k((xJ)_k)$ acts like $(n-r)$ terms
of interaction, encoding informations like the strength of the couplings, or
the nature of the clauses, etc.



The matrix $J$ describes how the variables interact.
In the more general case, the entries of $J$ are homomorphisms from $G$
to itself.
A specialization of this case is the one in which the entries of $J$ are in
a commutative ring with unit, $\mathcal{R}$,
and $G$ is an $\mathcal{R}$-module. A further specialization is the case in
which $G$ is
an arbitrary abelian group and $\mathcal{R}=\mathbb{Z}$.
In fact, every additive group is an
$\mathcal{R}$-module with $\mathbb{Z}$, where 
$n \cdot g = g+ \ldots + g$ ($n$ times).

An example of this case is a scalar field theory in which the $r$ variables
$\phi_i$ take a real value for each site $i$ of the lattice, and the
interaction terms depend on $(\phi_i-\phi_j)^2$, where $(i,j)$ is a bond of
the lattice.
Another example is a Gauge Potts Model, in which the $r$ variables $s_i$
take a value in $\mathbb{Z}_q$ for each bond $i$ of the lattice, and the
interaction terms depend on the linear combinations 
$\sum_{i \in \partial p} s_i$, where $p$ is a plaquette of the lattice.
In fact in both these models all the coefficients of $J$ are in 
$\{ 0, \pm 1\}$. In the first case, the entry $J_{ij}$ is non-zero 
when the site labeled by $i$
is an extremum of the edge labeled by $j$, while in the second case 
this happens when the edge labeled by $i$ is
on the border of the plaquette labeled by $j$, and the sign is determined by
the choice of orientation for edges and plaquettes.

Nonetheless, other mathematical structures can be considered. A case which is
not included in the situation above ($G$ is an $\mathcal{R}$-module) is when
the entries of $J$ are in a field $\mathbb{F}$,
and $G$ is a finite-dimensional vector space over $\mathbb{F}$ 
(say, $\mathbb{F}^{\,l}$).
The field $\mathbb{F}$ could be, for example, the real field
$\mathbb{R}$, the complex field $\mathbb{C}$, or a finite field $GF(q^k)$.

Moreover, we stress the fact that
the generality of our formulation does not require any underlying geometrical
structure for the system, and applies also to situations, typical of
disordered mean-field models, in which such a structure does not exist.



\section{Duality for product groups}
\label{prodsecbis}

\setcounter{footnote}{1}

Let us now repeat
the procedure of the previous section specialized to our
case, and write the corresponding duality transformation. The Boltzmann weight
is
\begin{equation}
\label{hamilt}
\exp \left( - \ham_{G;f;B}(x) \right)
=
\prod_{j=1}^{n} f_j((xB)_j)
\ef.
\end{equation}
The partition function is
\begin{equation}
\label{standardZ}
Z_{G;f;B}=
\int_{[x]}
\exp \left( - \ham_{G;f;B}(x) \right)
\ef{;}
\end{equation}
and, applying Fourier transform, can be restated as
\begin{align}
\nonumber
Z_{G;f;B} 
&=
\int_{[x]}
\prod_{i=1}^{r} 
f_i\left( x_i \right)
\!\!\!
\prod_{k=r+1}^{n} 
\!\!\!
f_{k}\left( (xJ)_k \right)
=
\int_{[x]}
\int^*_{[y]}
e^{i \vartheta_{G_2}(xJ, y)}
\prod_{i=1}^{r} 
f_i\left( x_i \right)
\!\!\!
\prod_{k=r+1}^{n} 
\!\!\!
\widehat{f}_{k}\left( y_k \right)
    \\
\nonumber
=&
\int^*_{[y]}
\int_{[x]}
e^{i \vartheta_{G_1}(-y J^T, x)}
\!\!\!
\prod_{k=r+1}^{n} 
\!\!\!
\widehat{f}_{k}\left( y_k \right)
\prod_{i=1}^{r} 
f_i\left( x_i \right)
=
\int^*_{[y]}
\!\!\!
\prod_{\phantom{xx}k=r+1}^{n} 
\!\!\!
\widehat{f}_{k}\left( y_k \right)
\prod_{i=1}^{r} 
\widehat{f}_{i}\left( (-yJ^T)_i \right)
\ef,
\end{align}
from which we deduce that
\begin{equation}
\label{dual}
Z_{G; f; B}= 
Z_{\widehat{G};\widehat{f};\widehat{B}}
\ef{,}
\end{equation}
where the dual $\widehat{B}$ of the matrix $B$ is 
intended exactly as in matroid theory
(see \cite{oxley}, chapt.~2).
\begin{equation}
B=
\bigg[ \ I_r \  \bigg| \ J \  \bigg]
\ef{,}
\qquad
\widehat{B}=
\bigg[ \, -J^T \  \bigg| \ I_{n-r} \, \bigg]
\ef{,}
\end{equation}
In the most general case of $J_{ij}\in \mathit{Hom}(G,G)$,
the quantities $(J^T)_{ji}$ 
are the dual homomorphisms $\widehat{J}_{ij}$
defined according to (\ref{dualhom}).

The duality interchanges the role of variables and interactions. Furthermore, if
the functions $f_j$ are almost constant (low coupling, or high temperature), 
the functions $\widehat{f}_j$ are very peaked (high coupling, or low
temperature), so the duality relates a regime of high (resp.~low) temperature
in the one-body terms with a regime of low (resp.~high) temperature in the
interaction terms, and vice versa.

\section{The coupling with gauge fields}
\label{qed_sec}

In this section we slightly generalize the formulation of the problem
introduced in sections \ref{gensec} and \ref{prodsec}, 
and then we show that, with a change of variables,
it can be recast in the original framework.

This analysis has one main reason:
in section \ref{qed_sec_bis} we will describe the
physical examples both of a scalar field theory, and of a discretized free
electrodynamic theory, and show that our duality holds for both systems. Then
we show that the duality holds also for the system in which electrodynamic is
coupled to the scalar field in the traditional gauge-covariant way.

We want to show that this is a particular case of a general
recipe, introduced in this section, and 
valid for the cases in which we introduce a set of gauge fields
coupled to the terms of interaction.
The change of variables corresponds to convert the
basis of scalar fields 
and gauge fields into the basis of fields and covariant
derivatives.

Given $2k$ abelian groups 
$G_1$, \ldots, $G_k$, $G'_1$, \ldots, $G'_k$,
we will consider the system defined by
\begin{itemize}
\item $k$ functions $f_i$ from the group $G_i$ to the field $\mathbb{C}$;
\item $k$ functions $\widehat{g}_i$ from the group $\widehat{G}'_i$
to the field $\mathbb{C}$;
\item $2k+1$ homomorphisms $\varphi_i$ between the groups $G_i$ and 
$\widehat{G}'_i$:
\begin{equation}
\xymatrix
{
0 \ar[r]^{\varphi_{-k}} &
G_k \ar[r]^{\varphi_{-(k-1)}} & \ldots \ar[r]^{\varphi_{-1}} & 
G_1 \ar[r]^{\varphi_0} & 
\widehat{G}'_1 \ar[r]^{\varphi_1} & 
\ldots \ar[r]^{\varphi_{k-1}} & \widehat{G}'_k
\ar[r]^{\varphi_{k}} & 0}
\end{equation}
\end{itemize}
and the Gibbs weight of a state $x=(x_1, \ldots, x_k) \in \bigotimes G_i$ is
\begin{equation}
\label{ham1pFactor}
\exp \left[ -\ham_{f,\widehat{g} ;\varphi_i}(x) \right]
=
\prod_{i=1}^{k} f_i(x_i - \varphi_{-i}(x_{i+1}))
\prod_{j=1}^{k} \widehat{g}_j (\varphi_{j-1} \cdots \varphi_{0} (x_1))
\ef;
\end{equation}
(it is understood that $x_{k+1}=0$ and $\varphi_{-k}(x_{k+1})=0$).

Consider the change of variables $y_i=x_i-\varphi_{-i}(x_{i+1})$,
with Jacobian 1.
If we introduce the $k \times k$ matrix of homomorphisms 
\begin{equation}
\psi_{ij}= 
\varphi_{-(i-1)} \varphi_{-(i-2)} \cdots \varphi_{j-1}
\ef,
\end{equation}
where the homomorphism $\psi_{ij}$ goes from $G_i$ to 
$\widehat{G}'_{j}$, we could restate the Gibbs weight above as
\begin{equation}
\label{ham1pbis}
\exp \left[ -\ham_{f,\widehat{g} ;\varphi_i}(y) \right]
=
\prod f_i(y_i) \prod \widehat{g}_j \Big( \sum_l \psi_{lj} (y_l) \Big)
\ef.
\end{equation}
As a matrix of homomorphisms from the set of groups $\{G_i\}$
to the set of groups $\{\widehat{G}'_j\}$ 
is a homomorphism
between the product groups $\bigotimes G_i$ and 
$\bigotimes \widehat{G}'_j$, we are in the conditions to apply
of the general procedure we have outlined in section \ref{gensec}.

The dual model interchanges the groups $G_i$ with the groups $G'_i$,
the functions $f_i$ with the functions $g_i$.
As the transposed homomorphisms are given by
\begin{equation}
(\psi^T)_{ij}=(\psi_{ji})^T = \varphi_{i-1}^T \cdots \varphi_{-(j-1)}^T
\ef,
\end{equation}
it corresponds to interchange the homomorphism
$\varphi_i$ with the transposal homomorphism $\varphi^T_{-i}$.
In this specialization of the general procedure, the relation
(\ref{dual1p}) would read:
\begin{equation}
Z_{f,\widehat{g} ;\{ \varphi_i \}}
=
Z_{g,\widehat{f} ;\{ \varphi^T_{-i} \}}
\ef.
\end{equation}
In the case we deal with product groups, as described in section 
\ref{prodsec}, the matrix $B$ will not be still in the standard form
representation of Matroid Theory when expressed in the $x$ basis. 
Nevertheless, it will be in a clear
block form with $k$ rows and $2k$ columns of blocks. 
If we call $J_i$ the linear applications corresponding to the
homomorphism $\varphi_i$, the block $B_{ij}$ would be given by
\begin{equation}
B_{ij} =
\left\{
\begin{array}{ll}
I & i=j\leq k \ef;
\\
-J_{-(k-i)} & i=j+1\leq k \ef;
\\
J_0 J_1 \cdots J_{j-k-1}
& i=k \ef; \quad j>k \ef;
\\
0 & \textrm{e.w.}
\end{array}
\right.
\end{equation}
The matrix $B$ can be easily diagonalized. This results in the change of basis
from $x$ variables to $y$ variables in the sense described above. 
The resulting matrix $B'$
will have blocks $B'_{ij}$ given by
\begin{equation}
B'_{ij} =
\left\{
\begin{array}{ll}
I & i=j\leq k \ef;
\\
0 & i \neq j \ef; j \leq k \ef;
\\
J_{-(i-1)} J_{-(i-2)}  \cdots J_{j-k-1} & j>k \ef.
\end{array}
\right.
\end{equation}
Now that $B'$ is in the standard form
representation, we can apply duality, and then, inverting the
diagonalization procedure described above, restate the dual problem in
the original block form. The transformation interchanges the
matrices $J_i$ with $J_{-i}^T$, 
with a minus sign on $J_0$, which goes into
$-J_0^T$.

\section{The solution of delta constraints}
\label{deltasec}

A general feature of our duality in the factorized case is that
the resolution of constraints taking the form of delta functions 
has a dual counterpart in the removal of trivial constant
factors, and vice versa.
This feature can be seen in a transparent way in a simple unstructured
problem, shown in section \ref{rcgm_sec}, but it is more
commonly used when the
duality is combined with other geometric properties of the systems,
as, for example, in the case of models defined over regular
finite-dimensional cell-complexes. Such a situation will be described
in section \ref{pmpgsec}.

Consider a system with $r$ variables and $n-r$ interactions, 
in the standard form described above, in the case
in which the first $r$ functions $f_i(x)$ are constant.
For simplicity, say that the last $(n-r)$ functions $f_k(y)$ are all
equal to a certain $f(y)$.
The model is described by the Gibbs weight
\begin{equation}
\label{hamiltM1}
\exp [-\ham(x)]
=
\prod_{j=r+1}^{n} g((xJ)_j)
\ef.
\end{equation}
Following the duality procedure, 
the dual Gibbs weight is given by
\begin{equation}
\label{hamiltM1D}
\exp [-\widehat{\ham}(\xi)]
=
\prod_{i=1}^{r} \delta((-\xi J^T)_i)
\prod_{j=r+1}^{n} \widehat{f}(\xi_j)
\ef.
\end{equation}
In the following, 
assume for simplicity that the incidence matrix $J$ has maximal rank.
\footnote{Matrices of non-maximal rank $r'<r$ 
would only imply trivial overall
  factors of the kind $|\mathbb{F}|^{(r-r')}$, or the necessity of fixing 
``by hands'' $(r-r')$ variables, as for example in planar Potts model, where a
factor $q$ originates from this fact, or in the XY-Model - Coulomb Gas
relation, in which we should avoid to sum over the variable associated to one
special site, in order to fix a ``ground'' for the potential.}
In this case it can be written in standard form $J=(I,J')$ by mean of a change
of variables, and a relabeling of the last $(n-r)$ indices. So we can rewrite
the Gibbs weights above as
\begin{gather}
\label{hamiltM1bis}
\exp [-\ham(x)]
=
\prod_{j=r+1}^{2r} f(x_j)
\prod_{j=2r+1}^{n} f((xJ')_j)
\ef;
\\
\exp [-\widehat{\ham}(\xi)]
=
\prod_{i=1}^{r} 
\delta(-\xi^{\textrm{dep}}_i- (\xi^{\textrm{ind}} {J'}^T)_i)
\prod_{j=r+1}^{2r} \widehat{f}(\xi^{\textrm{dep}}_j)
\prod_{j=2r+1}^{n} \widehat{f}(\xi^{\textrm{ind}}_j)
\ef;
\end{gather}
where in the last expression we already stressed that 
the variables $\xi^{\textrm{ind}}$ form an independent set, while
the variables $\xi^{\textrm{dep}}$ can be expressed in terms of them
when solving the delta constraints. In this case we obtain
\begin{equation}
\exp [-\widehat{\ham}(\xi)]
=
\prod_{j=r+1}^{2r} \widehat{f}((-\xi^{\textrm{ind}}{J'}^T)_j)
\prod_{j=2r+1}^{n} \widehat{f}(\xi^{\textrm{ind}}_j)
\ef.
\end{equation}
This is exactly the Gibbs weight obtained applying duality directly to the
weight in the (\ref{hamiltM1bis}), considered as a system with $r$ variables,
labeled with the indices
from $r+1$ to $2r$, and $n-2r$ interactions, labeled with the indices
from $2r+1$ to $n$.

In conclusion, we draw a scheme which resumes the procedure
described above:
\begin{equation*}
\begin{array}{ccc} 
\begin{array}{cccc}
\Big[\!\!\!\!& I, & I,\ J & \!\!\!\!\Big] \\ 
& 1 & \multicolumn{1}{c}{f}&
\end{array}
& \xrightarrow{\textrm{remove constant factors}} &
\begin{array}{cccc}
\Big[\!\!\!\!& I, & J &\!\!\!\! \Big] \\
& \multicolumn{2}{c}{f} &
\end{array}
\\
\xymatrix@M=0pt@R=15pt@C=90pt{{}
\ar@{<->}[dd]|*+<3pt>[F-:<5pt>]
{\scriptstyle{\textrm{Duality}}}\\ \\ {}}
&&
\xymatrix@M=0pt@R=15pt@C=90pt{{}
\ar@{<->}[dd]|*+<3pt>[F-:<5pt>]
{\scriptstyle{\textrm{Duality}}}\\ \\ {}}
\\
\begin{array}{cc}
\left[ \,
\begin{array}{c} -I \\ -J^T \end{array} \right. , 
\!\!\!&
\left. \begin{array}{cc} I & 0 \\ 0 & I \end{array} 
\, \right]
\\ \delta & \widehat{f}
\end{array}
& \xrightarrow{\textrm{solve the $\delta$}} &
\begin{array}{cccc}
\Big[ \!\!\!\!& -J^T, & I &\!\!\!\! \Big] \\ 
&\multicolumn{2}{c}{\widehat{f}}&
\end{array}
\end{array} 
\end{equation*}

\section{A sufficient condition for self-duality}
\label{selfdual_sec}

Duality is particularly useful whenever it is possible to recover that a
system, under suitable conditions, is self-dual.

In the following, let us denote by $X$
the instance of a problem (say, the lattice, the set of
couplings,\ldots), and by $\sigma$ a configuration of the
system (say, the values of all the spins in the system).

Duality means that there is an involution
$\{X\} \rightarrow \{X\}$ such that
\begin{equation}
Z[\widehat{X}]=f[X] Z[X]
\ef,
\end{equation}
with $f[X]$ an analytic function of $X$.

Of course a sufficient condition for an instance $X^*$ to be self-dual is
that it is a fixed point of the involution
\begin{equation}
\widehat{X_*}=X_*
\ef.
\end{equation}
If there is a group of invariances $\mathcal{U}$ such that
\begin{equation}
\forall \ u \in \mathcal{U} \qquad
Z[uX]=Z[X]
\ef,
\end{equation}
then a weaker condition for self-duality can be stated. It is sufficient that
some $u \in \mathcal{U}$ exists such that
\begin{equation}
u\widehat{X_*}=X_*
\ef.
\end{equation}
At this point, we make a short digression on disordered systems. For these
systems the variable $X$ describing the instance is itself a random variable, distributed
over some space $\{X\}$ with some normalized measure $\mu(X)$ given 
{\em a priori}.
We are interested in the thermodynamic quantities (that is,
thermalized over $\sigma$) of a system in a given instance $X$,
furtherly averaged over the 
distribution of disorder $\mu(X)$.
For example, the average free energy, up to $\beta$ factors, is given by
\[
F=\reval{ \log Z}= \int \dx \mu(X) \log Z[X]
\ef.
\]
Source variables can be introduced to calculate
the other thermodynamic quantities.
For example, to calculate the average magnetization we
can introduce a magnetic field $h$, and obtain
\[
F(h)=\reval{ \log Z(h)}\ef;
\qquad
M=\left. \dede{h} F(h) \right|_{h=0}
\]
similarly, to calculate the average two-point function 
$\reval{\eval{\sigma_i \sigma_j}}$ we can introduce a vector of local magnetic
fields $\{h_i\}$, one per site, and obtain
\[
\reval{\eval{\sigma_i \sigma_j}}
=
\int \dx \mu(X) \eval{\sigma_i \sigma_j}_X
=\left. \frac{\partial^2}{\partial h_i \partial h_j} F(h) \right|_{h=0}
\]
The action of duality on the space of instances $\{X\}$
naturally induces an action on
the space of normalized measures over $\{X\}$
\[
\widehat{\mu}(\widehat{X})=
\Big|
\frac{\partial{X}}
{\partial{\widehat{X}}}
\Big|
\mu(X(\widehat{X})) 
\ef.
\]
The condition for self-duality now applies to measures. The trivial condition
for a measure $\mu^*(X)$ to define a self-dual problem is
\[
\widehat{\mu^*}(X)=\mu^*(X)
\ef.
\]
Analogously to what we did above, if we have a group $\mathcal{U}$ of
invariances for the problem, we can make this condition weaker. If we have
a normalized Haar measure $\int_{\mathcal{U}}$ on the group, the condition
which exploits the symmetries of the problem is
\[
\int_{\mathcal{U}} \widehat{\mu^*}(u X)=
\int_{\mathcal{U}} \mu^*(u X)
\ef.
\]
Of course this condition can be fulfilled, by using the invariance of the Haar
measure, if for any fixed $X$ there is an element of the group $\mathcal{U}$
such that
\begin{equation}
\label{selfdualcond}
\widehat{\mu^*}(uX)=\mu^*(uv_X X)
\ef.
\end{equation}
In practice, in the case of a disordered system, we have a measure
on the space of $n$-uples of functions $\{f_i\}$, and of $r \times (n-r)$
incidence matrices $J$. In the case in which $r=n/2$, all the first $r$
functions $f_i$ are equal to a certain $f(x)$ and all the last $r$ functions
$f_k$ are equal to its Fourier transform $\widehat{f}(x)$, the dual set of
function is identical to the original one, and we only need to check
self-duality on the measure for the incidence matrix, $\mu(J)$.

Let the functions $f(x)$ and $\widehat{f}(x)$
be invariant under a subgroup of the
automorphisms of the space $G$,
$\mathcal{A} \subseteq \textrm{Aut}(G)$.

The elements $u$ of the group of invariance $\mathcal{U}$ are such that
$Z_{f;J}=Z_{f;uJ}$.

They are of the form
$u=(\pi; \{a_i\}, \{a_k\})$ with 
$\pi \in \mathfrak{S}_{\{1, \ldots, r\} } \times
\mathfrak{S}_{\{r+1, \ldots, n\} }$,
where $\mathfrak{S}_{\Lambda}$ is the permutation group on the set $\Lambda$,
and $a_i$, $a_k$ in $\mathcal{A}$, with $i=1, \ldots, r$ and
$k=r+1, \ldots, n$.
The action of $u$ on a matrix $J$ is
\[
(u J)_{ij}=a_j J_{\pi(i) \pi(j)} a_i
\ef,
\]
from which we have that a consistent
group composition is
\[
u' u= (\pi' \pi; \{a_{\pi'(i)} a'_i\},
\{a'_{k} a_{\pi'(k)}\})
\ef.
\]
The condition for self-duality (\ref{selfdualcond}) above, in our case can be
restated as the existence for each matrix $J$ of an element 
$u_J\in \mathcal{U}$ such that
\begin{equation}
\label{selfdualcondbis}
\mu^*(u_{J}(-J^T))=\mu^*(J)
\ef.
\end{equation}
Note that the Jacobian of the transform 
$J \rightarrow -J^T$ is one.
Exploiting the group $\mathcal{U}$ of invariances for the problem, we can
restrict our attention to measures which are central functions for the group 
$\mathcal{U}$. If we introduce the two-element group generated by the
transformation $\iota: J \rightarrow -J^T$, the self-dual measures are the
(fewer) central functions on the (larger) group 
$\mathcal{U} \rtimes \{ e, \iota \}$.

\section{An easy example: the Gaussian model}

As a first easy application of formula (\ref{dual}) we will consider 
a simple model exactly solvable via Gaussian integration.
Consider the system described by equation (\ref{hamilt}), 
specialized to the case in
which $G=\mathbb{R}$ and the functions $\{f_j\}$ are
\begin{subequations}
\label{effegauss}
\begin{align}
f_i(x) &=
e^{-x^2/2 \sigma_i^2} 
&
\mathrm{for}
& \quad i=1, \ldots, r
\ef{;}
\\ 
f_{r+k}(x) &=
e^{-x^2/2 \tau_k^2}
&
\mathrm{for}
& \quad k=1, \ldots, n-r
\ef{.}
\end{align}
\end{subequations}
We will introduce a notation {\em ad hoc} to label 
the partition function of this system:
\begin{equation}
Z_{\textrm{Gauss}}(J;\{\sigma_i\}, \{\tau_k\} )
\equiv
Z_{\mathbb{R}, B, \{f_j\}}
\ef{,}
\end{equation}
where the expressions (\ref{effegauss}) above for the $\{f_j\}$ and the
vectorial representation $B=[I_r|J]$ are understood.

We could rewrite equation
(\ref{hamilt}) in the more familiar form
\begin{equation}
\label{gwGM}
\exp( - \ham_{G; f; B}(\vett{x}))=
\prod_{j=1}^{n}
f_j((xB)_j)
=
e^{-\frac{1}{2}\vett{x}^{\dagger}
\left( S^{-2} + J T^{-2} J^T \right) \vett{x}}
\ef{,}
\end{equation}
where $S_{ii'}=\sigma_i \delta_{ii'}$
and $T_{kk'}=\tau_k \delta_{kk'}$. Using the convention of Table~\ref{t1}
for the measure $\int_{x}$ in the case $X=\mathbb{R}$ (that is, writing down
the proper number of $(2\pi)$ factors),
and applying the formula for Gaussian integration,
we have for the partition function
\begin{equation}
Z_{\textrm{Gauss}}(J;\{\sigma_i\}, \{\tau_k\} )
= \int \dx{x} \ 
e^{-\frac{1}{2} \vett{x}^{\dagger}
\left( S^{-2} + J T^{-2} J^T \right) \vett{x}}
=
\frac{(2 \pi)^{\frac{r}{2}} \det S}
{\sqrt{\det \left( I_r + M^T M  \right)}}
\ef{,}
\end{equation}
where we defined
\begin{equation}
M=
T^{-1} J^T S
\ef{.}
\end{equation}
From equation (\ref{dual})
we have that the dual partition function is
\begin{equation}
\label{ZrelGauss}
Z_{\textrm{Gauss}}(-J^T; \{\tau_k^{-1}\}, \{\sigma_i^{-1}\} )
=
\frac{(2 \pi)^{\frac{n-r}{2}}\det T^{-1}}
{(2 \pi)^{\frac{r}{2}}\det S}
\,
Z_{\textrm{Gauss}}(J;\{\sigma_i\}, \{\tau_k\} )
\ef{,}
\end{equation}
while, applying the formula for Gaussian integration on the dual formulation,
\begin{equation}
Z_{\textrm{Gauss}}(-J^T; \{\tau_k^{-1}\}, \{\sigma_i^{-1}\} )
= 
\int \dx{\vett{\eta}} \  
e^{-\frac{1}{2} \vett{\eta}^{\dagger}
\left( T^2 + J S^2 J^T \right) \vett{\eta}}
=
\frac{(2 \pi)^{\frac{n-r}{2}}
\det T^{-1}}{\sqrt{\det \left( I_{n-r} + M M^T  \right)}}
\ef{.}
\end{equation}
So the consequences of the duality can be restated in the algebraic equation
\begin{equation}
\det \left( I_r + M^T M \right)
=
\det \left( I_{n-r} + M M^T \right)
\ef{,}
\end{equation}
which is indeed true, although not totally trivial.
\footnote{
Consider the formal power series in $z$:
\begin{multline*}
\ln (\det (I-z M^T M))=\tr (\ln (I-z M^T M))=
-\sum_{n=1}^{+\infty} \frac{z^n}{n} \tr (M^T M)^n
\\
=-\sum_{n=1}^{+\infty} \frac{z^n}{n} \tr (M M^T)^n
=\tr (\ln (I-z M M^T))=
\ln (\det (I-z M M^T))
\ef{.}
\end{multline*}
}

\section{A randomly constrained Gaussian Model}
\label{rcgm_sec}

In this section we introduce a model in which some Gaussian distributed
variables in $\mathbb{R}^D$ are subjected to a set of random
constraints.

This model is a first example in which the matrix of the
homomorphisms $B$ really contains non-trivial group homomorphisms.
Furthermore, it exhibits in a clear way 
the general feature of our duality
described in section \ref{deltasec}: 
the resolution of delta constraints has a dual counterpart 
in the removal of trivial constant factors.

The model is defined as follows. Let $x_i$, $i=1, \ldots, r$, be
$D$-dimensional real vectors, and introduce the $(n-r)$ linear combinations
\begin{equation}
y_k^{\alpha}= \sum_{i=1}^r \sum_{\beta=1}^D 
x_i^{\beta} J_{ik}^{\beta \alpha}
\end{equation}
with $k=r+1, \ldots, n$, defined by the matrices $J_{ik}$. 
We will consider the case $2r>n$.

The Gibbs weight of a configuration is
\begin{equation}
\label{gwRCGM}
\dx x
\,
\exp [ -\ham_{J} (x) ]
=
\dx {}^D x_1 \cdots \dx {}^D x_r
\,
\exp \Big( -\frac{1}{2 \sigma^2} \sum_{i=1}^r x_i^2 \Big)
\cdot 
\prod_{k=r+1}^{n} \delta(y_k, 0)
\ef.
\end{equation}
Working in the spirit of disordered systems, we would have 
a probability measure over the quenched
disorder variables $J_{ik}$. Introducing the Haar
measure $\dx \mu_{D}$ over the group of special
orthogonal transformations $SO(D)$, a possible choice is to take the
variables $J_{ik}$ as i.i.d.~variables, distributed with $\dx \mu_D$:
\begin{equation}
\dx \mu(J)=
\prod_{
\substack{
i=1, \ldots, r
\\
k=r+1, \ldots, n}}
\dx \mu_D(J_{ik})
\ef.
\end{equation}
The problem has a large gauge group, which allows for the reduction
of the matrix $J$ in a standard vectorial form. 
We have mainly three class 
of manipulations on $B$ which do not 
modify the value of the partition function:
\begin{itemize}
\item permutations of the columns:
\[
J_{ik} 
\quad \longrightarrow \quad
J_{i\sigma(k)}
\qquad \sigma \in \mathfrak{S}_{n-r} 
\ef;
\]
\item global rotations on rows and columns:
\[
J_{ik}^{\beta \alpha}
\quad \longrightarrow \quad
R_{(i)}^{\beta \beta'} J_{ik}^{\beta' \alpha'}
R_{(k)}^{\alpha \alpha'}
\qquad
R_{(j)} \in SO(D), \ 1 \leq j \leq n
\ef;
\]
\item full Gauss algorithm on the rows:
\[
J_{ik}
\quad \longrightarrow \quad
C_{ij} J_{jk}
\qquad C
\in SO(r)
\ef.
\]
\end{itemize}
%
%
Thus, by change of variables we can put the matrix $J$ in the form
\[
J'=\begin{pmatrix} Y \\ -I_{n-r} \end{pmatrix}
\ef.
\]
Remark that this can be done always when the rank of $J$ is maximal, and this
happens with probability 1 for our measure $\dx \mu(J)$.

This procedure of diagonalization corresponds to the algebraic
resolution of the delta functions. In fact, the linear combinations are
\begin{equation}
(y')_k^{\alpha} = \sum_{i=1}^{r} \sum_{\beta=1}^D 
x_i^{\beta} (J')_{ik}^{\beta \alpha}
= 
\sum_{i=1}^{2r-n}
\sum_{\beta=1}^D 
x_i^{\beta} Y_{ik}^{\beta \alpha}
-x_{k-(n-r)}^{\alpha}
\ef,
\end{equation}
and, as the matrix $C$ above is orthogonal, the linear
transformation which relates the new $\{ y'_{k} \}$ 
to the old $\{ y_k \}$
has Jacobian equal to 1.
Defining
\begin{equation}
x_{k-(n-r)}^{\alpha}(x_{1}, \ldots, x_{2r-n}; Y)=
\sum_{i=1}^{2r-n}
x_i^{\beta} Y_{ik}^{\beta \alpha}
\ef,
\qquad
k=r+1, \ldots, n
\ef,
\end{equation}
and a transposal matrix $Y^T$ such that 
$(Y^T)_{ki}^{\alpha \beta}=Y_{ik}^{\beta \alpha}$,
the Gibbs weight (\ref{gwRCGM}) can be rewritten as
\begin{equation}
\label{gwRCGMbis}
\begin{split}
\exp [ -\ham_{J} (x) ]
& =
\exp 
\Big[ -\frac{1}{2 \sigma^2}  \Big( \sum_{i=1}^{2r-n} |x_i|^2
+ \sum_{j=2r-n+1}^{r} |x_{j}(x; Y)|^2 \Big)
\Big]
\\
& =
\exp 
\Big[ -\frac{1}{2 \sigma^2}
\sum_{i,j=1}^{2r-n}
x_i^{\alpha} (I + Y Y^T)_{ij}^{\alpha \beta} x_j^{\beta}
\Big]
\ef,
\end{split}
\end{equation}
which is of the form of the ``simple'' Gaussian model, equation 
(\ref{gwGM}), with $2r-n$ variables, $n-r$ constraints, 
and all the $\sigma_i$ and the $\tau_k$ equal to
$\sigma$, and the measure of integration $\dx x$ stands for 
$\dx {}^D  x_1 \cdots \dx {}^D  x_{2r-n}$.
So, with a slight abuse of notations, we can write 
the partition function of this model as
\begin{equation}
Z_{\textrm{Gauss}}(Y,\sigma)=
\int 
\dx x \ 
\exp \Big( -\frac{1}{2 \sigma^2}
x^{\dagger} (I_r + Y Y^T) x \Big)
\ef,
\end{equation}
%
We could apply the duality to this formulation of the
problem, using the general result of
equation (\ref{ZrelGauss}), obtaining
\begin{equation}
\label{ZrelRCGM}
Z_{\textrm{Gauss}}(Y,\sigma) = 
\frac{1}{(2 \pi)^{n-r}}
\left( \frac{2 \pi}{\sigma} \right)^{\frac{n}{2}}
Z_{\textrm{Gauss}}(-Y^T,\sigma^{-1})
\ef.
\end{equation}
We could also apply the general duality directly to the original
formulation. In direct formulation we have $r$ variables 
$x_1, \ldots, x_r$, 
with one-body functions $f_i(x_i)=e^{-x_i^2/(2 \sigma^2)}$, and
$(n-r)$ interactions, encoded in the delta functions
$g_k(y_k)=\delta(y_k,0)$. So that the matroid $B$ has the form
$B=(I_r, J)$.


Accordingly to the general theory, the dual problem has $(n-r)$
variables $\eta_1, \ldots, \eta_{n-r}$, with constant one-body functions 
$\widehat{g}_k (\eta_k) = \widehat{\delta}(\eta_k)= 1/(2 \pi)$,
and $r$ interactions given by
\begin{align*}
\widehat{f}_i(\xi_i)
&=\sqrt{\frac{2 \pi}{\sigma}}
\exp \left( -\frac{\sigma^2 \xi_i^2}{2} \right)
\ef;
&
\xi_i^{\beta} &=\sum_{k=r+1}^n 
\sum_{\alpha=1}^D
\eta_k^{\alpha} (-J^T)_{ki}^{\alpha \beta}
\ef.
\end{align*}
The dual matroid is given by $\widehat{B}=(-J^T, I_{n-r})$,
and the definition of the transposal matrix is 
$(-J^T)_{ki}^{\alpha \beta}=-J_{ik}^{\beta \alpha}$.
At this point, we can remove the identity block from the matroid, as
the functions associated to its columns are trivial constants. Then we
can diagonalize the matrix $-J^T$ using the ``transposal'' of the procedure
described above for $J$ (that is, moves on rows and columns are
replaced by the correspondent moves on columns and rows), and
obtain again
the partition function on the right side of (\ref{ZrelRCGM}).

We remark that, in this case in which the entries of the matroid 
$B$ are non-trivial homomorphisms, we can check directly that, 
accordingly with the general recipe of section \ref{prodsecbis},
the entries $\widehat{B}_{ki}$ of
the dual matroid are related to the {\em homomorphism-transposal} of the
entries $B_{ik}$ (that is, the internal indices $\alpha$ and $\beta$
are exchanged).
Moreover, in agreement with 
the general description of section \ref{deltasec},
we see how the resolution of delta constraints 
and the removal of constant factors are dual procedures.

\section{A generalized Potts-like model}
\label{genpottssec}

In this section we study the effects of the duality 
on a particular model which slightly generalizes 
the one we introduced in \cite{us}.
We deal with $r$ variables 
$x_i \in \{0, \ldots, q-1\}$
and a general
hamiltonian with $(n-r)$ 
Potts-like interaction terms provided by a matrix $J$
\begin{equation}
-
 \ham(x;\grid, \{K_k\}, \{h_i\}) = 
\sum_{k=1}^{n-r} 
K_k \delta \Big( \sum_i x_i \grid_{ik} \Big)
+
\sum_{i=1}^{r}
h_i \delta(x_i)
\ef{,}
\end{equation}
where the sums and the delta functions 
are intended on $\mathbb{Z}_q$, that is $x \equiv x + nq$.
The weights of the terms in the Hamiltonian are more conveniently expressed 
in terms of a vector $v\in \mathbb{R}^n$, defined as 
\begin{align*}
v_i(x) &=
e^{h_i}-1
&
\mathrm{for}
& \quad i=1, \ldots, r
\ef{;}
\\ 
v_{r+k}(x) &=
e^{K_k}-1
&
\mathrm{for}
& \quad k=1, \ldots, n-r
\ef{;}
\end{align*}
and of the functions
\begin{equation}
\label{effepotts}
f_j(x)=1+v_j \delta(x)
\ef{.}
\end{equation}
If we introduce the matroid $B=(I_r|J)$, we can reconduct the problem in the
general form (\ref{hamilt}):
\begin{equation}
\exp \Big( -\ham_{\mathbb{Z}_q, B, \{f_j\}}(x) \Big) = 
\prod_{j=1}^{n} 
\left[ 1+ v_j \delta ((xB)_j) \right]
\ef{.}
\end{equation}
We will introduce a notation {\em ad hoc} also for the 
partition function of this system:
\begin{equation}
Z_{\textrm{Potts}}(J;\{v_j\} )
\equiv
Z_{\mathbb{Z}_q, B, \{f_j\}}
\ef{,}
\end{equation}
where the expressions (\ref{effepotts}) above for the $\{f_j\}$
are understood.

From the form of the Fourier transform of $f(x)=1+v \delta(x)$:
\begin{equation}
\widehat{f}(\xi)=
\frac{v}{q} \left( 1 + \frac{q}{v} \delta(\xi) \right)
\ef{,}
\end{equation}
and with the convention of Table~\ref{t1}
for the measure $\int_{x}$ in the case $X=\mathbb{Z}_q$,
we have the relation between the partition functions
\begin{equation}
\label{zrelpotts}
Z_{\textrm{Potts}}(-J^T;\{q/v_j\} )
=
\frac{\prod_{i=1}^{r} v_i}
{\prod_{k=1}^{n-r} q/v_{r+k}}
Z_{\textrm{Potts}}(J;\{v_j\} )
\ef{.}
\end{equation}
Let us specialize the general duality transformation we have obtained 
to some simple cases, related to the model that we introduced in
Ref.~\cite{us}.
First of all, we will consider the case in which 
the magnetic field is constant
($v_j = v$ for all $j=1, \ldots, r$)
and all the couplings are equals
($v_j = w$ for all $j=r+1, \ldots, n$). 
The equation (\ref{zrelpotts}) in this case gives
\begin{equation}
\label{zrelpottsbis}
Z_{\textrm{Potts}}(-J^T; q/w, q/v )
=
\frac{v^r}
{(q/w)^{n-r}}
Z_{\textrm{Potts}}(J; v, w )
\ef{.}
\end{equation}
We can give a simple mathematical interpretation to
the limit $v \rightarrow 0$, $w \rightarrow \infty$ of the formula above.
In the two dual models, the partition functions, properly regularized, 
just count the
number of solutions of the linear equations $xJ=0$ and $y(-J^T)=0$.
In this limit case the duality relation states that the rank of
the matrix $J$ is equal to the rank of $-J^T$, which is a trivial fact
of elementary geometry.
This model is exactly solvable
for the ensemble of random matrices on a finite field, 
that is we know the exact
probability distribution for the free energy and other physical quantities,
and the duality relation
plays indeed a role in the solution of the problem \cite{us}.

A particular case of (\ref{zrelpottsbis}) is when only 
$v \rightarrow \infty$ (and then $w' \rightarrow 0$)
where again the partition functions should be properly regularized. The
explicit expression is
\begin{equation}
\sum_x \delta(Jx) (1+w)^{-\sum_i (1-\delta(x_i))}
=
\sum_{\eta} \prod_k 
\left[ \frac{w}{q} + \delta((-J^T y)_k) \right]
\ef{,}
\end{equation}
therefore, the duality relates a sum, 
restricted to the set of solutions of a linear
system $Jx=0$, where 
the difficulty of the problem is contained 
on the external field contribution,
to a sum, with no external field, on all the states of the spectrum, where the
difficulty is contained in the non-trivial correlations between the degeneracy
of the energy levels.
A quantitative study of this system will be done in a forthcoming paper
\cite{forthcom2}.

We can use the general strategy of section \ref{selfdual_sec} to investigate
some sufficient conditions on the measure $\mu_0(J)$ for equation 
(\ref{zrelpottsbis}) to relate the partition functions
of {\em the same} model at different values of the parameters $v$ and $w$.

As the weight functions involved are of the form ``delta + constant'', with no
disorder parameters, the subgroup of automorphisms 
$A \subseteq \textrm{Aut}(\mathbb{Z}_q)$ such that for a $a \in A$ we have
$f(a(x))=f(x)$ is isomorphic to
the classical group $\mathbb{Z}_q^*$.
\footnote{The group $\mathbb{Z}_q^*$ is defined as the integers in
$\{1, \ldots, q-1\}$ which are coprime with $q$; the group operation 
is the product, and the standard modulo $q$ identification is understood.}
That is, if $a$ is an
integer coprime with $q$, the corresponding automorphism on $\mathbb{Z}_q$ acts
as $a(x)=ax \mod{q}$.

The group of invariances obtained is sufficiently large that, following
the strategy described in section \ref{selfdual_sec},
the self-duality condition can be verified for almost every
reasonable ensemble of matrices.
For example, this is the case of random
matrices, of sparse matrices, both with non-zero elements chosen randomly
among $\{1, \ldots, q-1\}$ or all equal to $1$, 
and also of matrices with non-trivial
correlations between the entries, if the correlations are symmetric
under the exchanging of rows and columns, like the ensemble of incidence
matrices of random bipartite graphs with vertices of fixed connectivity.
An exception is the standard ensemble for the study of XOR-SAT problems
\cite{simplest}, which is essentially a particular case of our model, but the
matrices $J$ are constrained to have a fixed number of non-zero entries per
row, and do not have any constraint on the columns.

We note that, in the case of self-dual measure on the ensemble of matrices
$J$, and $v=w$, the self-dual coupling occurs for $v=\sqrt{q}$,
which is indeed the critical point also for the ordinary Potts model on a
geometrically self-dual planar graph. 
We will show in the next section how to recover explicitly the self-duality in
this case, in which other ``geometrical'' manipulations are required.

\section{Potts model on graphs and on planar graphs}
\label{pmpgsec}

The model discussed in the previous section is a generalization of ordinary
Potts model. In this section we will show how our duality relation can be
reconducted to the common duality of Potts model on planar graphs~\cite{dualpottsplan}.

Consider a connected graph $\Lambda$, 
at this stage not necessarily planar, with sites 
$s\in \mathcal{S}$, bonds 
$b\in \mathcal{B}$ and elementary cycles (or {\em loops})
$l\in \mathcal{L}$. Say that the number of sites, bonds and loops are
respectively $S$, $B$ and $L$.
If we consider the graph 
as embedded on an orientable surface,
the set $\mathcal{L}$ of the loops could be chosen to be the
set of the plaquettes 
(``local'' loops), plus the set of the generators of the first homology group,
that is, loops which turn around its handles.
We remark that, at this point, this division is not necessary.

We will give an arbitrary orientation to the bonds and to the loops. That is,
for each bond $b$,
$s_{\textrm{in}}(b)$ and $s_{\textrm{out}}(b)$ will be respectively the first
and second extremum of the bond. Furthermore, 
each loop $l$ will be associated with a
$B$-dimensional vector $J_b(l)$, such that, having arbitrarily
chosen an orientation for the loop, $J_b(l)$ is the number of times
that the loop $l$ is occupied by the bond $b$ in the same orientation,
minus the times it is occupied by $b$ in opposite orientation.

The Potts Model on the graph $\Lambda$ is defined by the set of variables
$\sigma_s \in
\{ 0, \ldots, q-1 \}
$ on the sites, and a set of interactions with coupling
constants $K_b$ on the bonds.
\footnote{We introduce an additive group
structure $\mathbb{Z}_q$ for the $q$ ``colours'' of the model, 
although ordinary Potts model Hamiltonian 
has a $\mathfrak{S}_q$ total permutational symmetry between the
colours.
We stress again the important fact that in general 
the symmetry group of the model {\em is not related} 
to the group introduced to perform the Fourier transform.}
The Hamiltonian is
\begin{equation}
\label{pghamilt}
-
 \ham_{\substack
{\textrm{Potts}\\\textrm{Graph}}}(\sigma; \Lambda, \{K_b \})=
\sum_{b \in \Lambda} K_b 
\, \delta(\sigma_{s_{\textrm{in}}(b)} - \sigma_{s_{\textrm{out}}(b)})
\ef{.}
\end{equation}
If we introduce the functions
\begin{subequations}
\label{effepg}
\begin{align}
\label{effepg1}
f_s(x) &= 1
\ef{;} & & \\
\label{effepg2}
f_b(x) &= 1 + v_b \, \delta(x)
\ef{;} &
v_b &= e^{K_b}-1
\ef{;}
\end{align}
\end{subequations}
and the matroid $B=(I_S|J_{sb})$, 
with 
$J$ a 
$S \times B$ matrix defined as
\begin{equation}
\label{matrixpg1}
J_{sb}=\left\{
\begin{array}{ll}
1 & s=s_{\textrm{in}}(b) \ef{;} \\
-1 & s=s_{\textrm{out}}(b) \ef{;} \\
0 & \textrm{e.w.}\ef{;}
\end{array}
\right.
\end{equation}
the partition function
\begin{equation}
\label{zpg1}
Z_{\mathbb{Z}_q, B, \{f_{s}, f_{b}\}}
=
\sum_{\sigma} \prod_s f_s(\sigma_s) 
\prod_b f_b \Big(\sum_s \sigma_s J^{(1)}_{sb} \Big)
\end{equation}
is in the general form (\ref{standardZ}).

Now we remove the useless identity functions $f_s(x)$, and make the change of
variables 
\begin{equation}
\label{chvar}
x_b(\sigma)= \sigma_{s_{\textrm{in}}(b)} -
\sigma_{s_{\textrm{out}}(b)}
\ef.
\end{equation} 
The weight of each configuration $\sigma$ is now just 
$\prod_b f_b(x_b(\sigma))$. Next, we want to perform the sum of the partition
function on the new variables $x$, instead that on the old $\sigma$. To this
aim, we need to know how many
configurations $\sigma$ correspond to a certain configuration $x$. The answer
is simple: they are exactly $q$ if the circuitation of $x$ on each loop is
zero (that is, $\sum_b J_b(l) x_b \equiv 0 \mod q$),
\footnote{More generally, for $\Lambda$ not connected,
we have a factor $q$ for each connected component.}
and zero if the circuitation is non-zero for some loop.
So we have
\begin{equation}
\label{chmeas}
\sum_{\sigma}
\quad \longrightarrow \quad
q \sum_x \prod_{l \in \mathcal{L}}
\delta \Big( \sum_b J_b(l) x_b \Big)
\ef{,}
\end{equation}
and, with the definitions
\begin{align}
f_b(x)&=1 + v_b \, \delta(x) \ef{;} &
f_l(x)&=\delta(x) \ef{;} \\
J_{bl} &= J_b(l) \ef{;} &
B' &= \Big(\, I_B
\, \Big| \, J_{bl}\, \Big) \ef{;}
\end{align}
we have again a system in the general form (\ref{standardZ}).
At this point we can perform the duality transformation. 
According to the general formula 
(\ref{dual}), and to the expressions for the Fourier transforms of 
$f_b(x)$ and $f_l(x)$, respectively
\begin{equation}
\label{ftgp}
\widehat{f}_b(\xi) = \frac{v_b}{q} 
\left( 1 + \frac{q}{v_b} \delta(\xi) \right)
\ef{;}
\qquad
\widehat{f}_l(\xi) = 1
\ef{;}
\qquad
\widehat{B'} = \Big(-J_{lb}
\, \Big| \, I_L \, \Big) \ef{;}
\end{equation}
the result of the transformation is
a system in the original general form (\ref{zpg1}).
If the matrix
$-J_{lb}$ is of the kind of $J_{sb}$, with exactly two non-zero
entries per column, a $+1$ and a $-1$, 
we can reconstruct a dual graphical version of the
original Potts model. 
As shown in detail in the \ref{app1}, this always 
happens when the original graph is planar.
In a few words, 
when a graph is planar, every bond appears in exactly two loops (the
plaquettes on the two sides). If we choose a uniform orientation for the
loops (all clockwise or all counter-clockwise), it appears in the two loops
with opposite sign. So, in this dual model, two neighbouring plaquettes
of the original graph appear to
interact with the dual coupling constant of the common bond
($v'_b=q/v_b$). If we want to reconstruct a graph such that 
the system has the variables on the sites, and
the interactions on the bonds, 
the desired
graph is exactly the dual graph of the
original one, in the sense of cell-complexes duality.

The procedure of this section can be applied with minor modifications to
obtain the $XY$-model~--~Coulomb Gas relation \cite{villain, nienhuis}, 
provided that the 
$XY$-model Hamiltonian is in the Villain form. Essentially, the only
difference is that the couple of dual groups is $(\mathbb{Z}_q, \mathbb{Z}_q)$
for the Potts Model, and $(U(1), \mathbb{Z})$ for the 
$XY$-model~--~Coulomb Gas relation.

\section{The coupling with gauge fields}
\label{qed_sec_bis}
\setcounter{footnote}{1}

In this section we describe the application of 
the procedure described in section \ref{qed_sec} to
a very well known physical situation. 
The abstract framework that we introduced should now assume a more transparent
meaning: that procedure mimics what happens if we 
introduce a new set of gauge fields, coupled with the terms of interaction, and then
a new term of the Hamiltonian depending from the physical content of these new
fields.

The paradigm of this procedure is the introduction of minimal electrodynamic interaction in a
scalar continuous field theory, with lagrangian
\begin{equation}
\mathcal{L}_0 = 
\int \!\! \dx \phi 
\left[ V(\phi) + \partial_{\mu} \phi \partial^{\mu} \phi \right]
\ef.
\end{equation}
The  derivative operator 
$\partial_{\mu}$
has to be modified into a covariant derivative operator $D_{\mu}$, defined by the action
on the fields as 
$D_{\mu} \phi= \partial_{\mu} \phi - A_{\mu}$. Then we have to add the dynamics of the gauge 
field,
$F_{\mu \nu} F^{\mu \nu}$, where the field tensor is defined as
$F_{\mu \nu} = \partial_{\mu} A_{\nu} - \partial_{\nu} A_{\mu}$.
The new lagrangian will be
\begin{equation}
\mathcal{L}_{\textrm{QED}} = 
\int \!\! \dx \phi \, \dx A_{\mu} 
\left[ V(\phi) + 
D_{\mu} \phi D^{\mu} \phi
+ \frac{1}{4 e^2} F_{\mu \nu} F^{\mu \nu} \right]
\ef{.}
\end{equation}
%
If we consider the statistical mechanic model whose Hamiltonian derives
from the lattice discretized
version of this model 
(as it is done, for example, in \cite{peskin}), we replace
partial derivatives $\partial_{\mu} \phi(x)$ with finite difference
operators 
$(\partial \phi)_{n; \mu} = \phi_{n+ \widehat{\mu}}-\phi_{n}$, 
so that, as usual, the
interactions lie on the bonds of the lattice. The linear combinations
$(\partial \phi)_{n; \mu}$ are the arguments of the interaction terms
in the Gibbs factor.

The discretized 
gauge fields $A_{n;\mu}$ couple to the terms of interactions, in such
a way that the finite difference of the fields, 
$(\partial \phi)_{n; \mu}$, are modified into the covariant derivative
$(D \phi)_{n; \mu}=(\partial \phi)_{n; \mu} - A_{n;\mu}$.

The discretized field tensor
$F_{n;\mu \nu}= A_{n;\mu} + A_{n+\widehat{\mu}; \nu }
-A_{n+\widehat{\nu}; \mu }-A_{n;\nu}$ takes the circuitation of the
vector field $A_{n;\mu}$ around each elementary plaquette
$(n; \mu \nu)$ of the lattice, and thus is a linear combination of
the entries $A_{n; \mu}$.

The current quadrivector
is given by the divergence of the electromagnetic tensor, 
$j_{n, \mu}=\sum_{\nu} (F_{n, \mu \nu} - F_{n- \nu, \mu \nu})$, and
its divergence, which is automatically zero, is given by
$(\nabla \cdot j)_n = \sum_{\mu} (j_{n, \mu} - j_{n- \mu, \mu})$.

\begin{figure}[t]
\label{fig1}
\begin{center}
\includegraphics[bb=122 493 491 697]{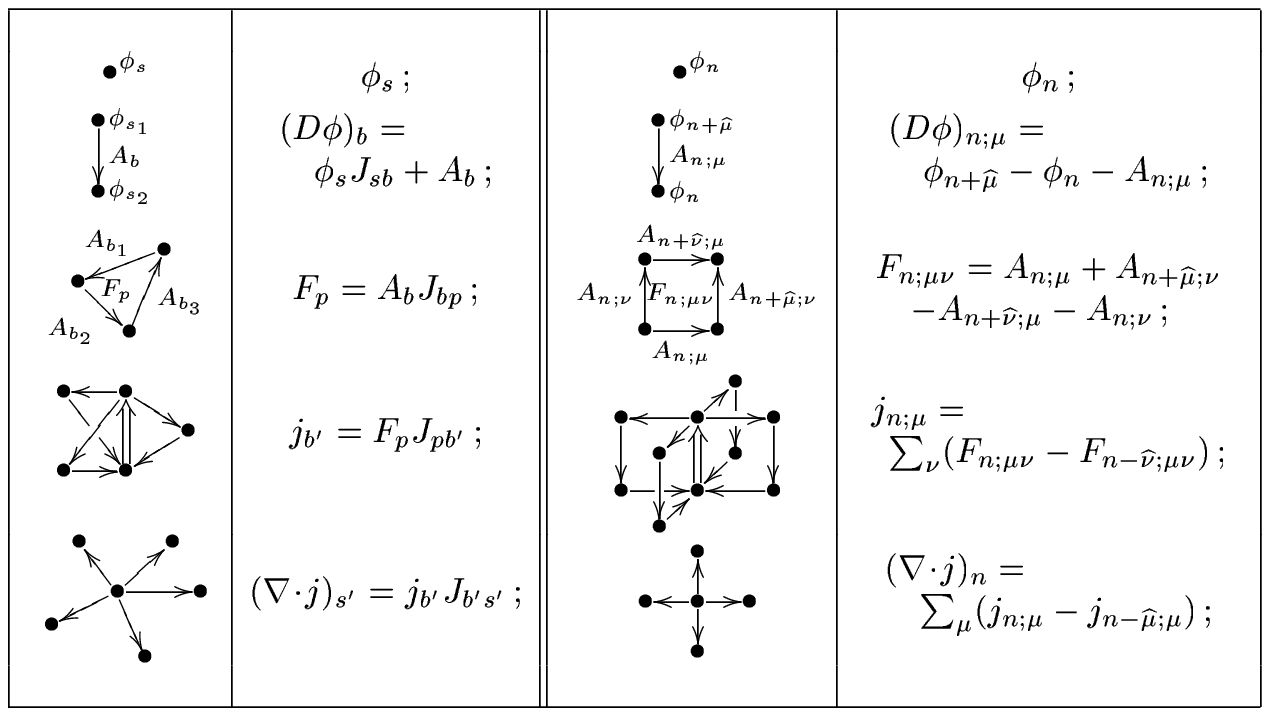} 
\end{center}
\caption{A table which collects the 
most
important quantities in discretized versions of QED, 
for regular lattices (right), and for
arbitrary graphs (left).}\label{f1}
\end{figure}

We can generalize this frame to an abstraction of the geometric
structure of a lattice, in which we have a set of sites $\{s\}$, a set
of bonds $\{b\}$, a set of plaquettes $\{p\}$ and a set of dual sites
$\{s'\}$. In the previous case we would have for the sites 
$\{s\}=\{n\}$, for the bonds $\{b\}=\{(n, \mu)\}$, for the
plaquettes $\{p\}=\{(n, \mu \nu)\}$,
and for the dual sites $\{s'\}=\{(n, \mu \nu \rho)\}$.

The action of gradient and
divergence operators is now performed via some adjacence matrices. In
particular we will introduce three matrices $J_{sb}$, $J_{bp}$ and
$J_{ps'}$, such that, given the two sets respectively of scalar fields
$\phi_s$ and of gauge fields $A_b$, we have
\begin{align}
(D \phi)_b &= \phi_s J_{sb} - A_b
\ef;
&
F_p &= A_b J_{bp}
\ef;
&
(\nabla \cdot j)_{s'}= F_p J_{ps'}
\ef.
\end{align}
We will be interested in the theory with Gibbs weight
\begin{equation}
\exp(-\ham_{\textrm{QED}}(\phi, A))=
\prod_s f_s(\phi_s) \prod_b f_b((D\phi)_b) 
\prod_p \widehat{g}_p (F_p) \prod_{s'}
\widehat{g}_{s'} ((\nabla \cdot j)_{s'})
\ef,
\end{equation}
with the four set of functions given by
\begin{subequations}
\begin{align}
f_s(x)&=1
\ef;
&
f_b(x) &= \exp \Big( -\frac{1}{2} x^2 \Big)
\ef;
\\
\widehat{g}_p (x) &= \exp \Big( -\frac{1}{4 e^2} x^2 \Big)
\ef;
&
\widehat{g}_{s'} (x) &= \delta(x)
\ef.
\end{align}
\end{subequations}
In our general frame of section \ref{prodsec} this model corresponds
to the matroid
\begin{equation*}
B= \begin{pmatrix}
I & -J_{sb} &&\\
& I & J_{bp} & J_{bp} J_{ps'}
\end{pmatrix}
\ef,
\end{equation*}
which is in the general form described in section
\ref{qed_sec}. Following the procedure described therein, we obtain
for the dual matroid
\begin{equation*}
\widehat{B}= \begin{pmatrix}
I & -J_{s'p} &&\\
& I & -J_{pb} & -J_{pb} J_{bs}
\end{pmatrix}
\ef,
\end{equation*}
in which sites are exchanged with dual sites, bonds are exchanged with
plaquettes, and the relative orientation of bonds and plaquettes is
reversed (in the geometrical duality for
three-dimensional lattices, this corresponds to the fact that
the orientations of bonds and dual bonds change in such a way that the
vector product changes parity).

Introducing the {\em ad hoc} notation for the partition function
\begin{equation}
Z_{\textrm{QED}}(e, B)=
\int \dx \phi_s \dx A_b
\exp \Big(
-\frac{1}{2} \sum_b (D \phi)_b^2 
-\frac{1}{4 e^2} \sum_p F_p^2 
\Big)
\prod_{s'} \delta((\nabla \cdot j)_{s'})
\ef,
\end{equation}
with the proper normalizations of the integrals, the duality relation
states that
\begin{equation}
Z_{\textrm{QED}}(e, B)
=
\frac{(2 \pi)^{\frac{r}{2}}}{(4 \pi e^2)^{\frac{n-r}{2}}}
Z_{\textrm{QED}}
\left( \frac{1}{2e}, \widehat{B} \right)
\ef,
\end{equation}
where, as always, $r$ is the rank of $B$, and in our case is equal to
the number of sites plus the number of bonds, and 
$(n-r)$ is the rank of $\widehat{B}$, and in our case is equal to
the number of dual sites plus the number of plaquettes.
Again we see how duality exchanges
a regime of high coupling with a regime of low coupling.

A concrete example could be the one of a cubic three-dimensional
lattice, with the ordinary notions of sites, bonds and plaquettes, and
the dual sites being the elementary cubic cells of the lattice. 
The geometric duality applied to the lattice naturally
exchanges sites with dual sites and bonds with plaquettes. In this
case, the presence of the delta function for the terms of current
divergence is redundant, as the constraint is automatically satisfied.

Variants in the spirit of disordered systems could involve a measure
$\mu(B)$ over an ensemble of random tessellations of the
three-dimensional space, or, more generally, over an ensemble in which
adjacency matrices are not directly derived from a geometrical
structure, and the condition $\nabla \cdot j = 0$ really needs to be
enforced via a delta function in the Gibbs weight.

In the case in which, following the recipe of section
\ref{selfdual_sec}, we have a self-dual measure $\mu^*(B)$, the fixed
point of the involution
for the electric charge $e$ is at the value $2 e^2 = 1$.

\section{Conclusions}

The notion of duality {\em 'a la} Kramers and Wannier, after its
original introduction in 1941, has been applied to a wide range of
periodic finite-dimensional systems of statistical mechanics. On the
contrary, small work has been done in the direction of a systematic 
description of duality for a class of models sufficiently large to
include, for example, mean-field disordered systems.

Our idea in this paper is that, in the Kramers and Wannier
formulation, a sort of two-step procedure is implicit: an 
{\em algebraic duality}, which exchanges variables and interactions,
and performs a Fourier transform on the terms of the Gibbs weight, and
a {\em geometric duality}, which, exploiting the geometrical structure
of the lattice, allows for a reinterpretation of the dual system in
a formulation which, in some contexts, is suitable for a direct
comparison with the original system.

In the more wide context of disordered systems the second step is
in general impossible. Nonetheless, in many cases the first step is
already suitable for a fruitful comparison of the two systems related
by duality.

The general recipe is the following: given some {\em variables} and
some {\em interactions}, valued on some abelian group $G$,
the Gibbs weight depends from two ingredients:
\begin{itemize}
\item two sets of functions, weighting
respectively the original variables, and the auxiliary variables
corresponding to some linear combinations of the original ones
(the {\em interactions});
\item the specific pattern of linear combinations of the variables, 
encoded in a matrix $J$ of group-homomorphisms.
\end{itemize}
The dual system exchanges variable-functions
and interaction-functions with the Fourier anti-transform and
transform of interaction-functions and variable-functions, and the new
pattern is given by the matrix $-J^T$ of transposed homomorphisms.

We give some concrete applications for specific simple models: a random Gaussian model, 
a random
Potts-like model, a random variant of discrete scalar QED\ldots~For
each of these systems, a short discussion over the duality relation is
proposed, and a hint over self-duality conditions is
suggested. Actually, in the cases of interest, this discussion is a
simple specialization of a general discussion done in section 8, and
analogous to the one already present in \cite{us}, where the duality,
in its first application, was called {\em variable-clause duality}.

An unexpected connection with Matroid Theory has emerged in the
analysis of the subject. The {\em a posteriori} justification for this
fact is that the pattern of interaction which is involved in the
definition of the Gibbs weight is intrinsic with respect to a choice
of basis for the space of configurations. Matroid Theory, in simple words, 
describes mathematical objects which convey basis-intrinsic
information on matrix-like objects, like the pattern of linear dependence
between the columns of a matrix.
Hence, it is small
surprise if the matroid related to the dual system is the matroid-dual
of the matroid related to the original system.

In a forthcoming paper we would discuss a further generalization of
this framework, in which the crucial concepts of 
{\em Fourier transform} and of {\em group-homomorphism} 
for abelian groups
are generalized to different integral transform and natural notions of
homomorphism, applied to mathematical structures different from
abelian groups. A first case is the one in which, for Partially Ordered
Sets, the natural notion of transform is the M\"obius transform, and
the natural notion of homomorphisms is related to the theory of Galois
connections \cite{forthmobius}.

                       \appendix

\section{Duality for graphical matroids and planar graphs}
\label{app1}

An important subset of regular matroids\footnote{That is, matroids which are
  representable over every field.} is the set of graphical matroids. Many
results exist on the subject (cfr., for example,
\cite{oxley}, chapt.~5), derived in the spirit of generality characteristic of
Matroid Theory. In this appendix we will show some constructive procedures
which apply to particular classes of graphical matroids, so in this sense they
are less general, but more clear restatements of known facts.
Hopefully, they will be useful to
understand how to interpret the classical formulations of dualities for SM
models on planar lattices inside our general setting of the duality.

A precise terminology will be introduced in the body of the section. Here we
resume the main results:
\begin{theor}
The ``site-bond'' matroid 
$J_{sb}$ 
and the ``loop-bond'' matroid 
$J_{lb}$ 
of a
graph $\mathcal{G}$ are dual matroids.
\end{theor}
\begin{theor}
The ``site-bond'' matrix 
${J'}_{sb}$ 
of a planar graph $\mathcal{G}$ is equal to the
``loop-bond'' matrix 
${J'}_{lb}$ 
of the corresponding
geometrical dual graph $\mathcal{G}^*$, and conversely the
``loop-bond'' matrix 
of $\mathcal{G}$ is equal to the
``site-bond'' matrix
of $\mathcal{G}^*$.
\end{theor}


\subsection{Matroid duality for graphical matroids}

Consider a
graph $\mathcal{G}$, with vertices (or sites) 
$s \in V$ and edges (or bonds) $b \in E$.
We give an arbitrary orientation to the bonds. That is,
for each bond $b$,
$s_{\textrm{in}}(b)$ and $s_{\textrm{out}}(b)$ will be respectively the first
and second extremum of the bond.
\begin{equation}
\xymatrix@M=-1pt@C=18pt{
\bullet \ar[r]^>{b}^(-0.5){s_{\textrm{in}}(b)} & 
\ar@{-}[r]^(1.5){s_{\textrm{out}}(b)} & \bullet}
\end{equation}
Define a 0-form $\varphi$ as a function from $V$ to a field $\mathbb{F}$, 
and a 1-form $\psi$ as a function from $E$ to $\mathbb{F}$. The operators
$\partial$ and $\partial^*$ transform 0-forms in 1-forms and vice versa with
the following definition:
\begin{align}
\partial \varphi(b)
& = \varphi(s_{\textrm{in}}(b))- \varphi(s_{\textrm{out}}(b)) \ef;
\\
\partial^* \psi(s)
& = \sum_b \psi(b) 
(\delta_{s, s_{\textrm{in}}(b)} - \delta_{s, s_{\textrm{out}}(b)})
\ef.
\end{align}
Intuitively, they correspond to the discretization of the 
gradient operator on scalar fields and to
the divergence operator on vector fields, generalized to arbitrary lattices.

The sites are implicitly a particular case of 0-form,
$\varphi_{(s)}(s')=\delta_{s,s'}$. The set of 0-forms 
is a vector space
over $\mathbb{F}$, 
of dimension $|V|$.
It is naturally decomposed in two vector spaces, $\mathcal{S}$ and
$\mathcal{C}$, where 0-forms in $\mathcal{C}$ are constant on each of
the $C$ connected components of $\mathcal{G}$ (so
$\dim(\mathcal{C})=C$), and 0-forms in $\mathcal{S}$ are obtained by
quotient (so $\dim(\mathcal{S})=S=|V|-C$). Projection on
$\mathcal{S}$ is obtained via the operator 
$\frac{1}{2} \partial^* \partial$.

Analogously, the set of 1-forms is a vector
space $\mathcal{B}$ over $\mathbb{F}$, of dimension $B=|E|$.
The bonds are a particular case of generic 1-forms, 
where it is understood $\psi_{(b)}(b')=\delta_{b,b'}$,
and are a basis for the space $\mathcal{B}$.
We will define the loops as the 1-forms $\psi$ such that 
$\partial^* \psi=0$. Elementary plaquettes are particular cases of
loops: if $\partial p$ is the border of an oriented plaquette $p$, 
$\psi_{(p)}(b)$ is the characteristic function on $\partial p$.
The set of loops is a vector space $\mathcal{L}$ over $\mathbb{F}$, 
of dimension $L$.

Note that $S+L=B$ as a consequence of Euler formula.
More deeply, the vector space $\mathcal{B}$ is decomposed into
$\partial \mathcal{S} + \mathcal{L}$, that is, a generic 1-form $\psi$
is uniquely decomposed into $\psi = \partial \varphi + \chi$, with
$\varphi \in \mathcal{S}$ and $\chi \in \mathcal{L}$ 
(Hodge decomposition).


The patterns of linear dependence of the set of bonds projected onto the 
vector spaces 
$\partial \mathcal{S}$ and $\mathcal{L}$
are matroids. They will be called respectively the {\em site-bond} and the
{\em loop-bond} matroids.
Given a choice of basis for $\mathcal{S}$ and $\mathcal{L}$, 
we will denote the
corresponding vectorial representations as $J_{sb}$ and $J_{lb}$.
That is, for a set of 0-forms $\{ \varphi_s \}$ 
chosen as a basis for $\mathcal{S}$, the entries of $J_{sb}$ will be
\begin{equation}
J_{sb} = \partial \varphi_s(b)
\ef,
\end{equation}
and for 
a set of 1-forms $\{ \psi_l \}$ 
chosen as a basis for $\mathcal{L}$, the entries of $J_{lb}$ will be
\begin{equation}
J_{lb} = \psi_l(b)
\ef.
\end{equation}
We will give a particularly skill choice of basis, for which the two matroids
are both in standard vectorial form, and manifestly dual.

Choose a ``ground'' site $s_0 \in V$, and a spanning tree 
$\mathcal{T} \subseteq \mathcal{G}$. Change the orientation of the bonds such
that for each site $s \neq s_0$ there is exactly one out-going bond 
in $E(\mathcal{T})$. Label this bond $b_s$.
Say that $s \succ s'$ iff an oriented path $\gamma_{s' \rightarrow s}$ in
$\mathcal{T}$ exists. Choose a basis for $\mathcal{S}$ as
\begin{equation}
\varphi_s(s')= \left\{
\begin{array}{ll}
+1 & s \succ s' \\
0 & \textrm{o.w.}
\end{array}
\right.
\end{equation}
So we have
\begin{equation}
\partial \varphi_s(b_{s'})= \delta_{s, s'}
\ef;
\qquad
\partial \varphi_s(b_l)= \left\{
\begin{array}{ll}
+1 & s \succ s_{\textrm{in}}(b_l), \ s \nsucc s_{\textrm{out}}(b_l) \\
-1 & s \succ s_{\textrm{out}}(b_l), \ s \nsucc s_{\textrm{in}}(b_l) \\
0 & \textrm{o.w.}
\end{array}
\right.
\end{equation}
We will choose a basis for $\mathcal{L}$ such that for each bond 
$b \in E \setminus E(\mathcal{T})$ there is exactly one 1-form $\psi_l$ 
such that $\psi_l(b)$ is non-zero. 
So we can label the edges in 
$E \setminus E(\mathcal{T})$ as $b_l$. 
More precisely, the 1-form $\psi_l$ corresponds to the only self-avoiding
closed path on $\mathcal{T} \cup \{ b_l \}$, oriented according to $b_l$. 
Analytically this can be expressed as
\begin{equation}
\psi_l(b_{l'})= \delta_{l,l'}
\ef;
\qquad
\psi_l(b_s)= \left\{
\begin{array}{ll}
-1 & s \succ s_{\textrm{in}}(b_l), \ s \nsucc s_{\textrm{out}}(b_l) \\
+1 & s \succ s_{\textrm{out}}(b_l), \ s \nsucc s_{\textrm{in}}(b_l) \\
0 & \textrm{o.w.}
\end{array}
\right.
\end{equation}
It is immediate to check that $J_{sb}$ and $J_{lb}$ are in standard vectorial
form, respectively with the identity block on the left and on the right.
From the observation that $\partial \varphi_s(b_l)=-\psi_l(b_s)$ we obtain
that they are dual in matroid sense:
\begin{equation}
J_{sb}=(I,Y_{sl})\ef; \qquad 
J_{lb}=(-Y^T_{ls},I) \ef.
\end{equation}
As we have found an explicit choice of basis for which the standard form of
the site-bond matroid and of the loop-bond matroid are dual, we have
implicitly proved that the matroids are dual, which is the claim of Theorem~1.

\subsection{Graphical duality for planar graphs}

In the previous section we have seen how the ``abstract''
site-bond and the loop-bond matroids for a given graph are dual in the sense
of matroid duality. This is true in general, as the matroid definition is
intrinsic, and specifically we have verified it in a particular choice of
basis, for which both the matroids are 
in standard form.

In this section we show the result of Theorem~2, that the 
site-bond and the loop-bond incidence matrices for a connected 
planar graph and for its
graphical dual are exchanged by the graphical duality. This result relies on a
particular choice for the loops as the elementary plaquettes of the lattice,
which is the more natural for the case
of planar graphs, but in this sense is more accidental. 

A graph $\mathcal{G}$ is said to be {\em planar} when it is embeddable on a
genus-$0$ surface with no crossings between the bonds. Consider the
case in which $\mathcal{G}$ is connected.
The classical Euler relation for
this kind of graphs claims that $V+F=B+2$, where $V$ is the number of sites,
$B$ is the number of bonds and $F$ is the number of faces.

One can immediately define the site-bond 
and loop-bond incidence matrices, 
$A_{sb}$
and 
$A_{lb}$, for
non-oriented bonds and loops. Just state that
$A_{sb}$ is 1 if the site $s$ 
is an extremum of the bond $b$ and 0 elsewhere, and
$A_{lb}$ is 1 if the bond $b$ is a side of the face 
$l$ and 0 elsewhere.

The graphical dual $\mathcal{G}^*$ is defined by the following procedure.
Replace each plaquette $l$ by a dual vertex $s^*$ 
(white bullets in Fig.~\ref{f1}), 
each bond $b$ by a dual (dashed) bond $b^*$, which crosses it and
connects the dual vertices of the two neighbouring plaquettes,
and each site $s$ with the dual plaquette $l^*$ delimited
by the dual bonds of the bonds outgoing from the site.
It is clear that
\begin{equation}
\label{Asbl}
A_{s^*b^*}(\mathcal{G}^*)=A_{lb}(\mathcal{G})
\ef;
\qquad
A_{l^*b^*}(\mathcal{G}^*)=A_{sb}(\mathcal{G})
\ef.
\end{equation}
\begin{figure}[t]
\label{fig_plangraph}
\begin{center}
\includegraphics[bb=135 592 461 693]{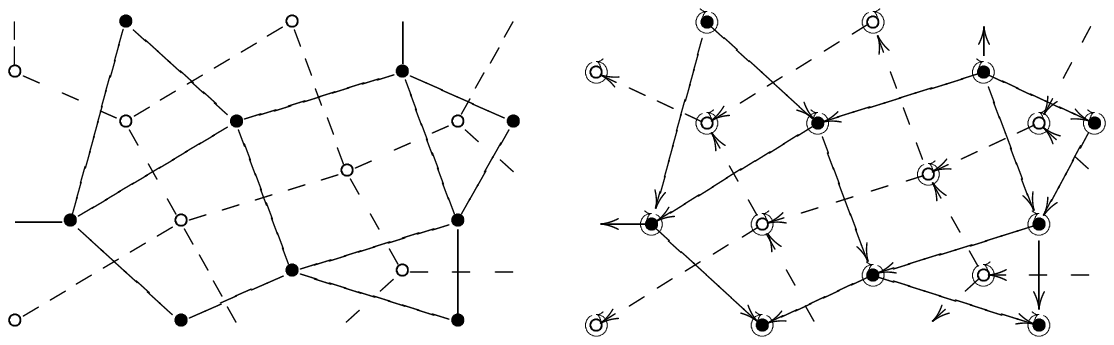} 
\end{center}
\caption{Graphical dual for a planar graph (left), 
and a consistent choice of orientations for the bonds 
and directions of rotations for the faces for the relation (\ref{Jsbl}) to
hold (right).}
\end{figure}
If we put an orientation on the bonds and a direction of rotation 
on the faces, the matrices 
$A_{sb}$
and 
$A_{lb}$ 
will be promoted to some matrices
$J'_{sb}$
and 
$J'_{lb}$, in which some minus signs appear.
Again because of the planarity of the graph, we could give a recipe such that
the relation (\ref{Asbl}) above is promoted to a relation
\begin{equation}
\label{Jsbl}
J'_{s^*b^*}(\mathcal{G}^*)=J'_{lb}(\mathcal{G})
\ef;
\qquad
J'_{l^*b^*}(\mathcal{G}^*)=J'_{sb}(\mathcal{G})
\ef.
\end{equation}
The recipe is the following: choose the direction of rotation to be the same
for all the faces,
(for example clockwise),
and opposite between faces and dual faces.
Choose the orientation of the bonds arbitrarily. The orientation of the dual
bonds must be such that the vector product between each bond and its dual bond
has the same sign with respect to the normal versor outgoing from the surface.

Both the matrices 
$J'_{sb}$
and
$J'_{lb}$ 
are not of maximal rank, and the only linear relation
is given by the sum of all the rows with equal coefficients.
This accounts
for the offset of 2 in the Euler formula $V+F=B+2$, in fact the number of
``independent sites'' in the loop-bond matroid (0-forms up to an
arbitrary constant)
is $S=V-1$, and the number of independent loops is
$L=F-1$, from which we recover the relation $S+L=B$.

                  \ack 

We thank Alan Sokal for introducing us to the relevance of the theory of
matroids for Potts-like systems, and for valuable discussions in the reading
of the manuscript.

              \references  

\end{document}